\documentclass[conference]{IEEEtran}
\IEEEoverridecommandlockouts
\usepackage{cite}
\usepackage{amsmath,amssymb,amsfonts}
\usepackage{algorithmic}
\usepackage{graphicx}
\usepackage{textcomp}
\usepackage{xcolor}
\def\BibTeX{{\rm B\kern-.05em{\sc i\kern-.025em b}\kern-.08em
    T\kern-.1667em\lower.7ex\hbox{E}\kern-.125emX}}
\begin{document}

\title{Accuracy vs Performance: An abstraction model for deadline constrained offloading at the mobile-edge\\
{\footnotesize}
\thanks{This work is funded by Research Ireland projects 18/CRT/6222 and 13/RC/2077\_P2}
}

\author{\IEEEauthorblockN{Jamie Cotter, Ignacio Casti\~{n}eiras, Victor Cionca}
\IEEEauthorblockA{\textit{Dept. of Computer Science} \\
\textit{Munster Technological University}\\
Cork, Ireland \\
jamie.cotter@mycit.ie, ignacio.castineiras@mtu.ie, victor.cionca@mtu.ie}
}

\maketitle

\begin{abstract}
\label{sec:abstract}

In this paper, we present a solution for low-latency deadline-constrained DNN offloading on mobile edge devices. 
We design a scheduling algorithm with lightweight network state representation, considering device availability, communication on the network link, priority-aware pre-emption, and task deadlines. 
The scheduling algorithm aims to reduce latency by designing a resource availability representation, as well as a network discretisation and a dynamic bandwidth estimation mechanism. 
We implement the scheduling algorithm into a system composed of four Raspberry Pi 2
(model Bs) mobile edge devices, sampling a waste classification conveyor belt at a set frame rate. 
The system is evaluated and compared to a previous approach of ours, which was proven to outcompete work-stealers and a non-pre-emption based scheduling heuristic under the aforementioned waste classification scenario.
Our findings show the novel lower latency abstraction models yield better performance under high-volume workloads, with the dynamic bandwidth estimation assisting the task placement while, ultimately, increasing task throughput in times of resource scarcity. 

\end{abstract}


\section{Introduction}
\label{sec:introduction}

Mobile devices are tackling increasingly complex tasks such as image, video, and audio processing, using Machine Learning and AI models. Although capable of running these locally, devices can become overwhelmed if the task rate increases. In that case a solution is to offload tasks to nearby idle devices \cite{distream}. Some applications can be time critical (e.g. self-driving cars), which means tasks have deadline constraints.


The processing time of task scheduling algorithms adds to the task latency, with the main contributor being the search through all network resources for a valid allocation of the task. There is a trade-off between the accuracy of the scheduler and its time performance. This paper contributes to this accuracy vs. performance trade-off by developing a data structure that makes scheduling and reserving resources faster. Unlike existing approaches, such as calendar queues \cite{calendarqueue}, we represent computational resources as a series of overlapping time windows. However, rather than simply queueing tasks, we instead reserve computational resources by representing them as periods of availability that tasks subtract from.

Specifically, in this paper, we present a solution for low-latency task scheduling by generating a lightweight representation of network resources (both computational and communication), which is used by our scheduler for deadline-constrained DNN offloading. The scheduler supports prioritisation of tasks, and high-priority tasks can preempt low-priority ones to meet their deadline.
Our lightweight representation is composed of two data-structures: a discretisation of the network link and the representation of computational resources as guaranteed periods of availability. Both are designed to reduce the latency overhead perceived by tasks that request network resources, ultimately increasing task throughput in times of resource scarcity.

The system is evaluated on Raspberry Pi 2 (model Bs) and compared to our prior work, which was proven to outcompete workstealers and a non-pre-emption based scheduling heuristic in the context of a waste classification application. Based on the results of our experiments, our findings are the following:
\begin{itemize}
    \item Lower latency abstraction models yield better performance under high-volume workloads, but the loss in accuracy results in poorer performance under lighter workloads.
    \item Dynamic bandwidth estimation mechanisms assist the system in avoiding erroneous task placement; however, they do not offset the performance hit from network congestion.
    \item The rate at which bandwidth estimation tests are performed can have a significant impact on system performance. High rates result in a loss in performance through the introduction of a high amount of congestion in network communication and internal system performance because the associated data structures must be regenerated.
\end{itemize}
The structure of the paper is as follows: Section \ref{sec:related_work} presents related work on both light-weight mobile edge scheduling algorithms and network discretisation approaches. Then, Section \ref{sec:system_design} presents the design of the system. Section \ref{sec:algorithms} discusses the three main factors in our approach: resource availability representation, network discretization, and their integration into the scheduling algorithm. Next, Section \ref{sec:implementation} presents the implementation of the experiment. The results are discussed in Section \ref{sec:results}. Finally, Section \ref{sec:conclusions} provides some conclusions.

\section{Related Work}
\label{sec:related_work}

\textbf{Offloading DNN inference} The offloading of DNN inference tasks to nearby devices has been extensively studied. Preliminary works such as \cite{deepthings}, \cite{adaptiveparallel}, or \cite{hadidi} propose data and model partitioning of a single inference task and distributing between devices. This results in a larger total processing time than processing on a single host, due to large communication overheads \cite{jamiepimrc}. More recent works such as Distream \cite{distream} and Jellyfish \cite{jellyfish} focus on offloading entire DNNs, for scenarios where the client must process multiple tasks at once, e.g., classifying of multiple targets in the same frame.

\textbf{Resource scheduling} Allocating and scheduling resource usage for tasks in a cluster is a well-known problem, with mature open source solutions such as Apache YARN \cite{yarn}. Most cluster managers aim to maximise the job throughput (\textit{makespan}), with load-balancing, or tail distribution as secondary objectives. Some solutions (Borg \cite{borg}, Natjam \cite{natjam}) differentiate production from best-effort tasks, using pre-emption of the latter to free up resources for the former. Unlike our work, most cluster managers do not enforce task deadlines. Doing so requires \textit{resource reservation}, with the issue highlighted by Rayon \cite{rayon}, who propose a resource definition language and a MILP solver, as well as heuristics for handling generic reservation of resources. Rayon is closest to our work. The types of resource considered vary in the literature, with most solutions focusing on computation (CPU and memory). Tetris \cite{tetris} is one of the few solutions that highlights the importance of considering network and storage resources, which are subject to congestion and therefore can delay jobs. Oktopus \cite{oktopus} and \cite{proteus} are network bandwidth reservation solutions for multitenant data centres. Oktopus considers static bandwidth demands, while Proteus enables jobs whose requirements vary with time.

\textbf{Data structures for reservation} Another distinction between our work and cluster managers is in the management of time: existing works use equal-length time slots, whereas we consider variable time windows, which significantly complicates resource allocation. Calendar queues \cite{calendarqueue} are an $O(1)$ data structure for queuing timestamped events, which has been used in scheduling computing tasks, e.g. packet processing \cite{programmable_calendar_queue}. Our solution is different because it also handles reservation of resources, not just queuing.

\section{System Design}
\label{sec:system_design}

\begin{figure}[h]
    \centering
    \includegraphics[width=\columnwidth]{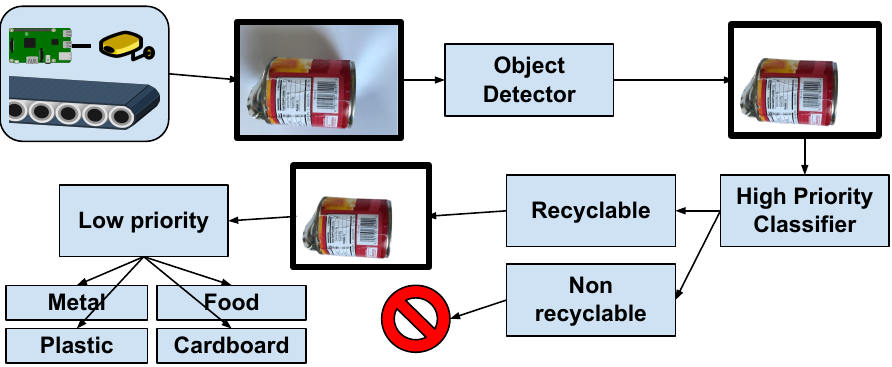}
    \caption{Task Pipeline}
    \label{fig:task_pipeline}
\end{figure}

This work considers the waste classification system from our previous work \cite{deadlinepreempt}, where a set of edge devices monitor waste items that pass down conveyor belts running at a set speed, with the goal of classifying waste into recyclable classes. We consider the task pipeline from our previous work shown in Fig. \ref{fig:task_pipeline}. It consists of three stages: Stage 1 is an object detector that identifies whether waste is present on the conveyor belt in a given frame; Stage 2 is a low-complexity binary classifier that determines if the waste is recyclable or non-recyclable; and finally, if recyclable waste is detected, the pipeline advances to Stage 3, where a high-complexity classifier categorises the frame into one of four classes of recyclable waste.

The random distribution of waste items across the conveyor belts leads to varying workloads over time (e.g., at a given point in time a device might be handling several waste items while another device is idle). Computational offloading allows the system to take advantage of the inconsistent nature of the workload by offloading excess work to idle devices and achieve higher system throughput.

The system operates in a centralised manner, where scheduling decisions are made by a controller that maintains the state of communication and computation resources based on information received from the edge devices.

\subsection{Network Resources}
Unlike existing work, we consider the network as a resource that must be allocated, whereas most computational schedulers (except, e.g., Tetris \cite{tetris}) only consider compute resources. However, the literature shows the importance of the network, which can introduce congestion that will delay the start time of offloaded tasks, leading to deadline violations. To address this, we model the network link such that each slot represents a discrete time-window that allows us to convert a time-stamp and immediately receive a viable communication window which greatly reduces the search space. Additionally, we consider computational availability as a series of continuous time interval windows of variable duration, instead of fixed-size time slots. This makes scheduling easier, as querying available resources is now a containment query instead of an overlapping range search, which requires checking through all existing tasks.

The throughput of the network link may vary throughout the system execution, as it is affected by background traffic. The system uses a dynamic bandwidth estimation mechanism that allows the controller to schedule tasks while respecting the variation in network throughput. 


\section{Data structures and Algorithms}
\label{sec:algorithms}
In this paper, we present an abstraction model for fast lightweight scheduling decisions for high volume task offloading in homogeneous mobile edge networks. In our previous work, we used a basic representation of the network state, the devices contained allocated tasks, and the network link contained allocated communication windows. This representation meant that it was quick to maintain as each insertion (when tasks are allocated) and each removal (when tasks are preempted, violate their deadlines, or complete), both scaling linearly with the number of tasks. However, querying such data-structures is slow, as the available capacity of computational and communication resources must be calculated each time by performing an overlapping range search, which increases the latency tasks experience. Unlike our previous work, we instead utilise a resource model of computational resources that represents guaranteed periods of resource availability tailored to each application type. Additionally, we discretise the time-horizon of the network link to allow quicker query times. Whilst operations to maintain the state of these data structures are slower, when implemented into our scheduling algorithm, we are able to reduce the search space for task placement in both computation and communication, allowing quicker query times and, in turn, reducing the latency a task perceives when requesting resources.

\subsection{Data Structures}
\subsubsection{Resource Availability Model}

\begin{figure}
    \centering
    \includegraphics[width=\columnwidth]{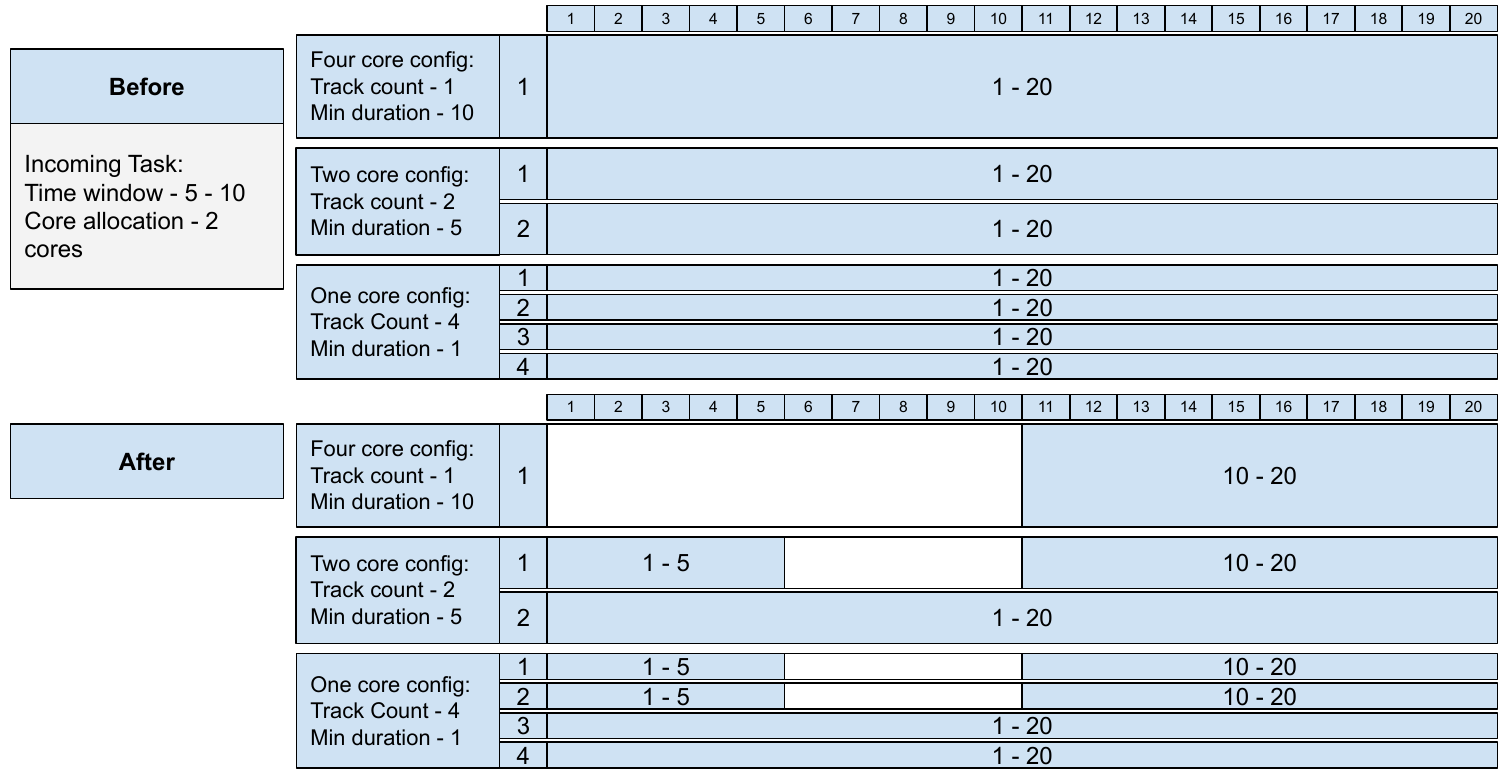}
    \caption{Example of writing to a resource availability data structure} 
    \label{fig:res_avail_dia}
\end{figure}

In order to implement a low-latency scheduling algorithm, we implement an abstraction model that transforms the workload of devices from a set of tasks assigned to the device to a series of available windows. In doing so, a query for the capacity of a device is no longer an overlapping range-based query that must iterate over the entire workload of a device to calculate the total resource usage that overlaps with a tasks desired window. Instead, with the new model based on available windows, the query where the first window that contains our desired time-slot allows for an early exit of the search function.

Furthermore, since each device has $n$ number of cores and the task configuration associated with them has $j$ number of cores required, we generate $n / j$ number of tracks representing resource availability (i.e., if a task requires two cores and the device has four, then it can process two tasks simultaneously). As seen in Fig. \ref{fig:res_avail_dia}, each entry in a resource availability list is sized at a minimum of the processing time and capacity required for a task, ensuring that the first window found will accommodate the task. We have three different types of application configurations with their own processing durations and resource requirements; therefore, each device must maintain an individual resource availability list for each application configuration. 
Every availability window in a resource availability list contains two parameters $t1$ and $t2$, which are the two time points of the time window it represents. The resource availability lists as a whole have three parameters associated with them, minimum core capacity, minimum duration and track count.

After a window has been selected within a resource availability list, it must be bisected with the desired time slot creating from 0 to 2 new windows (left-hand side, right-hand side), depending on how the desired time slot is contained within the selected availability slot. These new windows must satisfy the minimum core and duration requirements of the resource availability list if they are to be inserted. 

The trade-off of this approach is that, whilst it allows for quicker query times during task scheduling, the write process to record task allocation (which occurs after a task has been allocated resources) is more expensive, as each task allocated must be written across each availability list for the device (not just the list for their application configuration). However, write operations are treated as background operations, performed after task allocation, ensuring that there is no latency penalty perceived by allocated tasks.

Task pre-emption has a higher computational cost. The resources of the preempted task must be released. We cannot reinsert a reclaimed set of resources into the list, as each window within the list only represents that the given time window satisfies the minimum core requirements rather than total usage. Therefore, we reconstruct the entire set of resource availability lists for a given device from the active workload of the device. 
To do so, for each task configuration, we first construct a fully available resource availability list and then begin to bisect the resource availability lists and remove segments for each existing task assigned to the device in the same fashion that task allocation is performed.

\subsubsection{Network Link Discretisation}
\begin{figure}
    \centering
    \includegraphics[width=\columnwidth]{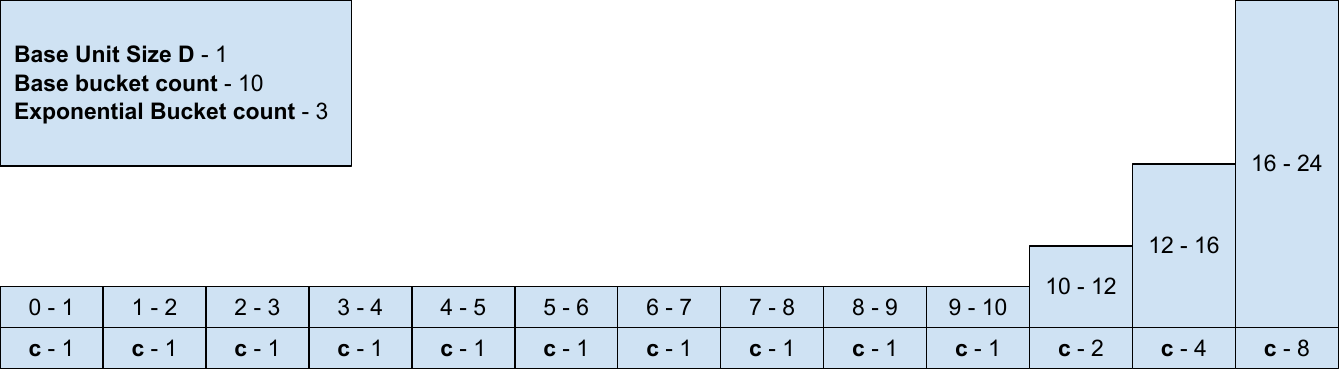}
    \caption{An example of the discretised network link} 
    \label{fig:net_link_image}
\end{figure}

In order to implement a quick network scheduling solution, we discretise the network link. To do so, we only schedule the dominant factor in communications on the network link. In the context of our work, this is the image transfer performed when a task is offloaded from one device to another. We first take the maximum image size possible for our image classification model and calculate the average transfer time based on an estimated bandwidth value derived from network performance tests. This serves as the size of the base unit of transfer $D$ of the elements in the discretised link. 

As shown in Fig. \ref{fig:net_link_image}, each element $b$ is a bucket that may contain $c$ number of communication tasks that we treat as the bucket capacity. Each bucket $b_i$ represents a time window composed of two time points $[t_1^i, t_2^i)$. The time window a bucket represents is $t_1^i (\equiv t_2^{i-1})$ and $t_2^i (\equiv t_1^i + (c^i \cdot D))$

The construction of the network link representation involves the following steps. Take the current time point $t_p$ and round it up to the nearest multiple of $D$, this forms the starting point of the network link discretisation, which will be referred to as the current time of reasoning $t_r$. We create the first $n$ number of buckets with a capacity of $1D$ to provide higher accuracy for potential windows in the near future. However, to address memory limitations and reduce the size of the discretisation, after $n$ number of buckets, we create $j$ number of buckets that have an exponentially increasing capacity size and corresponding time length.

We can obtain an index to query the discretisation by using the current time point $p$ with the following formula to obtain a base index $base\_index = ((t_p - t_r) + (D - ((t_p - t_r) \% D))) / D$. If the base index is less than the number of base buckets, then we return $floor(base\_index)$ otherwise we return $floor(log2(base\_index) + 2)$.
Once we have obtained an index, we then check to see if the bucket is at full capacity; if it is, we iterate forward until we find a bucket that can accommodate an additional communication task.

The network link discretisation must be reconstructed whenever the system internal network bandwidth estimate is updated. As the new network link discretisation begins at a further point in the time horizon than in the previous instance, we must downshift all existing items in the old network link to appropriate buckets in the new network link using a cascade function. To do so, we iterate over the old network link and for each bucket and each item in each bucket, applying the network link query function to retrieve an index in the new link. If the returned index is a negative value, then the communication task has already been completed and will be excluded; otherwise, we insert the communication task into its new index.

\subsection{Scheduling Algorithms}
\subsubsection{High-Priority Task Scheduling}
Our high-priority tasks are executed locally on their source device due to tight deadline constraints. Therefore, the high-priority scheduling algorithm does not concern itself with scheduling on the network link. 
To allocate a high-priority task we must calculate the processing time window it will occupy in the network, we first obtain the current time point $t_p$ that forms $t_1$, we then add the estimated processing duration to $t_1$ to form $t_2$. Using this time window, we perform a containment query on the High-Priority resource availability list specific to the source device. If a window is found, we allocate the task and perform the write operation to all of the availability lists associated with the device. Otherwise, we generate a pre-emption request for the source device at the calculated time window.

\subsubsection{Low-Priority Task Scheduling}
Unlike the high-priority task allocation algorithm, which allocates resources for a single task within a request, the low-priority task allocation algorithm attempts to allocate resources for $n$ number of tasks within a DNN scheduling request. Our low-priority tasks can be executed in two configurations, a two-core configuration that runs slower and a faster four-core configuration. The system uses a conservative resource allocation approach, with a preference to allocate two cores to a task, only allocating four cores if a two core allocation will violate task deadlines.
Therefore, we first take the current time point $t_p$ and the task deadline $d$ to see which resource configuration is viable (or, if neither will satisfy the deadline, exit early). This resource configuration also informs the resource availability list that will be queried (i.e. a two-core configuration will query the two core resource availability list). Next, we find a potential communication slot for each task within the request (not all of these slots will necessarily be used, as a task may find placement on its source device). Using the current time $t_p$ as $t_1$ and the task deadline as $t_2$ we form an estimated placement window for our task. We then use this placement window to perform a multi-containment query that returns each window that contains our desired placement window performed in parallel across each device in the network. If the number of windows returned is less than the number of tasks, then we cannot satisfy the request and exit. Otherwise, we prioritise windows from the source device and, to ensure that offloaded tasks are balanced across the network, we shuffle the remote devices and begin cycling through the devices taking one window at a time until each task has a valid processing window. While that is happening, the resource availability and network discretisation structures are updated to reflect the new allocations.

\subsubsection{Pre-emption Scheduling}
When a high-priority task cannot find placement, it issues a pre-emption request for the device it was generated from in the time window in which it attempted to find placement. To perform pre-emption, we iterate over the list of tasks assigned to the device, selecting a low-priority task that overlaps with its time window. If multiple low-priority tasks overlap with the time window, then we select the task with the farthest deadline for pre-emption. However, the resource availability model does not allow for a window to be reinserted as the total capacity is unknown. Therefore, for the device where pre-emption is occurring, we create new empty resource availability lists for each configuration and perform write operations based on the existing workload on the device. When a low-priority task is preempted, it will have a chance to receive reallocation by entering the low-priority scheduling algorithm once again.

\section{Implementation}
\label{sec:implementation}
We implement our system on Raspberry Pi 2 Model Bs that act as image processors and a 2020 Macbook Pro (M1) that runs the controller. The edge devices and the controller communicate over Wi-Fi using 802.11n. The controller runs the scheduling algorithms which is implemented in C++ 17 (compiled using clang 19.1.5); the edge devices run inference managers implemented in Python 3.11.4. The DNN tasks are processed using TensorFlow lite \footnote{https://www.tensorflow.org/lite} and our DNN model is based on YoloV2. 

The network discretisation relies on network throughput estimation. At the start of the experiment, the controller initially performs an iperf3 test with each edge device to generate a baseline estimate of network throughput. The initial estimate will be updated periodically (in our main experiments, this parameter is set to 30 seconds). First, the controller selects an edge device at random to host the bandwidth test. This device then sends 10 1400 byte pings to each other edge device in the network, measures the round-trip time of each ping, and uses it to calculate the bits per second of each ping. These values are returned to the controller which uses an exponentially weighted moving average (EMWA) (where an alpha of 0.3 is used) to adjust the internal bandwidth estimation before triggering a reconstruction of the network discretisation. 


Each task and task configuration (i.e. high-priority, low-priority with two-cores and four-cores) have fixed processing times based upon our benchmark tests, which result in fixed-length time slots for each stage and configuration type. The values are as follows: high-priority (0.98s), low-priority two-core (16.862s), and low-priority four-core (11.611s).
To minimise the impact of system load and of hardware variations during run-time on our low-priority tasks, we use the standard deviation from benchmark tests as padding on the processing time.

For the experiments, each edge device also runs a secondary Python application acting as our experiment manager. It is responsible for generating the frames for the pipeline at regular intervals (corresponding to conveyor belt speed), as well as for determining the processing deadlines of the pipeline. If the controller can successfully allocate a high-priority task, then its execution is simulated by having the experiment manager sleep for the allotted window.
If a high-priority task is determined to have spawned a set of low-priority tasks, it issues a low-priority request to the controller that can contain from 1 to 4 DNN tasks. If the controller can allocate the resources required by these tasks, it forwards their allocation to the inference manager of the chosen hosts.

For the purposes of our experiment setup, we use the same input image for each DNN task. DNNs require inputs to be of a certain size. In real-world scenarios, the waste items of our application would be extracted from the source frame and resized before being sent to the DNN.

To evaluate the various levels of task load, we model the experiment using trace files. These configurations control the distribution of high-priority and low-priority tasks (DNNs) generated per frame. Each entry in a trace file represents the workload for four devices in a given frame. Here, a device in a frame can have one of the following values: -1 (no object is detected), 0 (a high-priority task is generated but with no low-priority request afterward), and 1..4 (a high-priority task is generated and a low-priority request with n number of DNN tasks is generated after it completes). In our experiments, we use five different trace files representing different distributions of generated DNN tasks: in uniform devices, we generate 1..4 tasks with equal probability; in weighted X (x in 1..4) devices, we predominantly generate X tasks, with the network load increasing with X.

A new task pipeline is generated every ~18.86 seconds. We derive this value from running this system with the minimum viable completion time for the object detector, a high-priority task, and one low-priority DNN task partitioned across two cores.


\section{Results}
\label{sec:results}
To evaluate the impact that performance has over accuracy in deadline-constrained scheduling, we focus on several key metrics related to the scheduling latency and to the task completion rate of the systems we are analysing. The main metric that we focus on is the frame completion rate, as the main utility of the system is whether or not it can process image classification tasks. To fully understand the relationship between performance and accuracy, we look at the completion rate of the system under various levels of network load. We also examine the average latency incurred under these scenarios for both high and low-priority tasks, in order to see if latency increases with load and if it significantly impacts task completion rate.

We perform several evaluations of the systems outlined. First, we perform a comparative analysis of our proposed system utilising a scheduling abstraction model against a more exhaustive and accurate scheduler with higher latency overheads under several weighted loads. Then we examine the performance of our proposed system when we vary the frequency of the dynamic bandwidth update mechanism. Finally, we examine the effectiveness of our proposed system and the bandwidth mechanism under variable network conditions by artificially inducing network congestion using a traffic generator.

\begin{table}[]
\begin{tabular}{l|l}
\textbf{Experiment}                                                                   & \textbf{Label} \\ \hline
\begin{tabular}[c]{@{}l@{}}Weighted N (1 .. 4) - Pre-emption \\ Scheduler\end{tabular} & WPS\_N         \\ \hline
Resource Availability Scheduler Weighted N (1 .. 4)                                   & RAS\_N         \\ \hline
Bandwidth Interval Tests N (1.5, 5, 10, 20, 30) seconds                        & BIT\_N         \\ \hline
\end{tabular}
\\
\caption{Graph Legend}
\label{tab:graph_legend}
\end{table}

\begin{figure}
    \centering
    \includegraphics[width=\columnwidth]{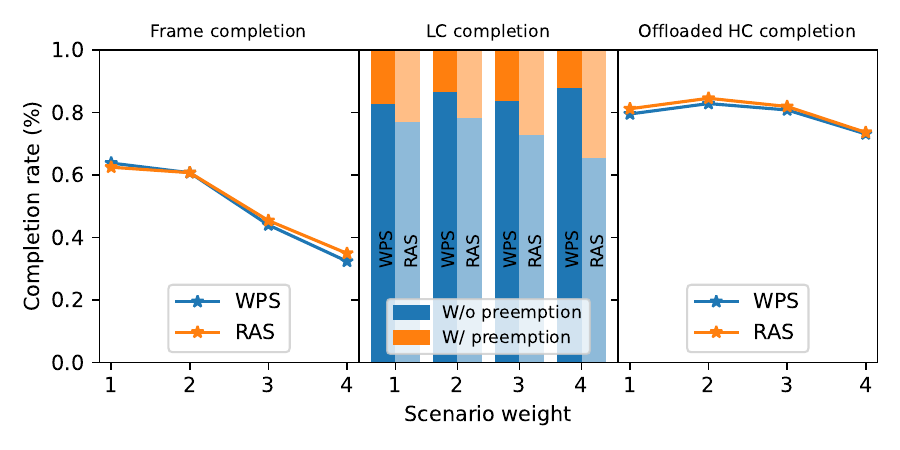}
    \caption{Task Completion across various categories.} 
    \label{fig:task_var_completion}
\end{figure}

\subsection{Accuracy vs Performance}
To measure the frame completion rate, we consider a frame completed if its high-priority task and all its low-priority tasks are completed (irrespectively of how many 0..4 low-priority tasks it generates). We observe in Fig. \ref{fig:task_var_completion} that under the lightest load, the more exhaustive approach taken in WPS yields better results. This is in part due to the fact that our proposed system makes certain trade-offs in accuracy that will not be necessary under a lighter uncongested network. However, we see that, as soon as the network becomes weighted toward generating two tasks, performance equalises between the two systems (our proposed system completes one more frame). Then, our proposed system begins to significantly outperform the WPS system at a weighted load of 3 and this gap grows in weighted load of 4. At weighted 3 the network is more likely to generate a load of tasks too great to process locally (as our devices have four cores, they can process at most two DNN tasks with a two-core allocation locally); therefore, both systems are forced to offload more frequently. This indicates two things: when tasks are allocated to their local device, there is no need to occupy any resources on the network link, reducing the search space for network communication slots; however, when tasks are offloaded (therefore WPS can reap the advantage of higher accuracy without paying the price of increased search time), the occupied link slots increase search times for subsequent task allocation requests.

\begin{figure}
    \centering
    \includegraphics[width=\columnwidth]{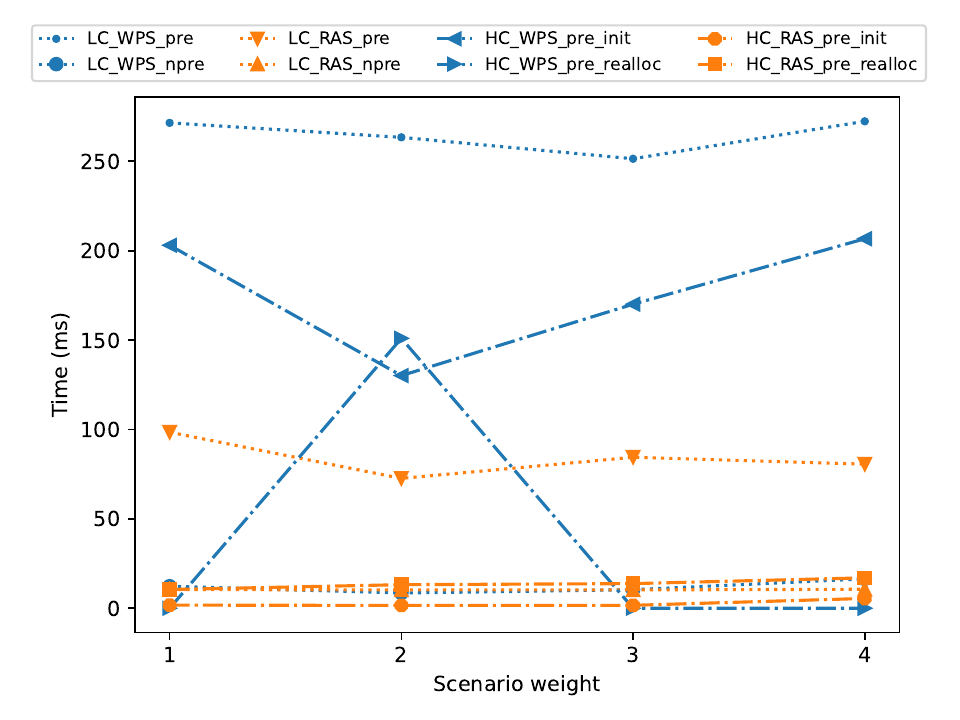}
    \caption{Scheduling latency by initial allocation and pre-emption/reallocation scenarios for both schedulers.} 
    \label{fig:task_var_latency}
\end{figure}

The impact of latency on the performance of the system can be seen in figures \ref{fig:task_var_latency} and \ref{fig:task_var_completion}, showing the average latency of high-priority tasks and their completion, with and without pre-emption. The WPS system in all scenarios is able to allocate more tasks without invoking pre-emption, indicating a better task-placement mechanism. While the latency incurred for allocating a high-priority task without pre-emption in both the WPS system and our proposed solution remains below 15ms across all loads, our proposed solution experiences a much lower latency in pre-emption scenarios, as the WPS never drops below 250ms in the pre-emption scenario whilst our proposed solution never exceeds 100ms. The impact of this is further seen in figures \ref{fig:task_var_latency} and \ref{fig:task_var_completion}, which show the average latency of low-priority task allocation with and without reallocation, as well as a breakdown of the completion of low-priority tasks. Across all scenarios, the WPS system incurs much higher initial allocation latency than our proposed solution, with its lowest result under the W2 scenario at 140ms and its highest at 205ms. On the other hand, our proposed solution remains under 6ms across all scenarios. The WPS system very rarely manages to successfully reallocate tasks following. However, when it does it experiences an average latency of 150ms or 10ms more than its initial allocation for the weighted load. On the other hand, our proposed solution successfully reallocates tasks under every weighted load with an average latency of 10ms, 13ms, 14ms and 17ms respectively. We see this in the high-priority completion rate, where our proposed solution successfully reallocates 217, 175, 186, and 236 tasks across the four weighted loads, respectively. 
Although the WPS scheduler can allocate more tasks overall, the latency in the high-priority pre-emption scenario has a knock-on effect on reallocation. As reallocation can only begin once the high-priority task has completed pre-emption, the longer it takes, the closer to the preempted tasks deadline reallocation begins. Furthermore, reallocating the task itself has significant latency overheads in the WPS scenario; the reduction in latency can be seen as the difference between the sum of the high-priority and low-priority pre-emption and reallocation latencies in the WPS and RAS scenarios.
Additionally, we can see that the WPS scheduler frequently has a much higher number of tasks that violate their deadlines, even if the total number of low-priority tasks completed is higher. This indicates that the impact latency has on system performance is significant, as any task that violates its deadline also invalidates the entire frame it belongs to. The RAS system has a lower number of tasks allocated, but they are more likely to complete. The benefits of the network link representation of the RAS system are also observed in Fig. \ref{fig:task_var_completion}, which shows the completion rate of offloaded low-priority tasks. While the WPS completes more low-priority tasks overall, when observing the completion of offloaded tasks, the difference between the two systems diminishes. This further reinforces that one of the benefits of the RAS system is related to its bandwidth mechanism and its network representation.

\begin{figure}
    \centering
    \includegraphics[width=\columnwidth]{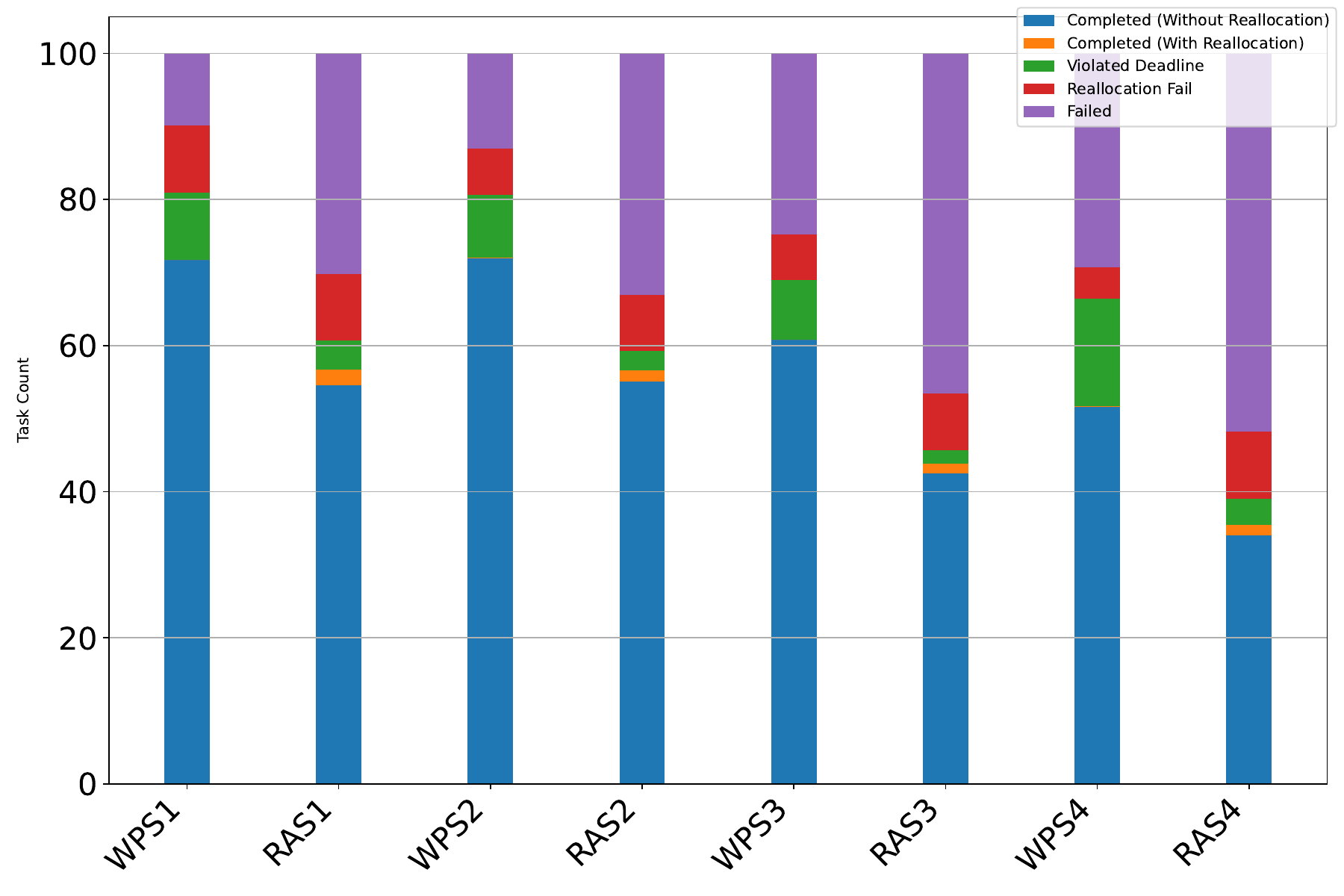}
    \caption{Low-priority high-complexity completion by mechanism.} 
    \label{fig:low_priority_high_comp_res}
\end{figure}

\subsection{Bandwidth Interval Rate}
\begin{figure}
    \centering
    \includegraphics[width=\columnwidth]{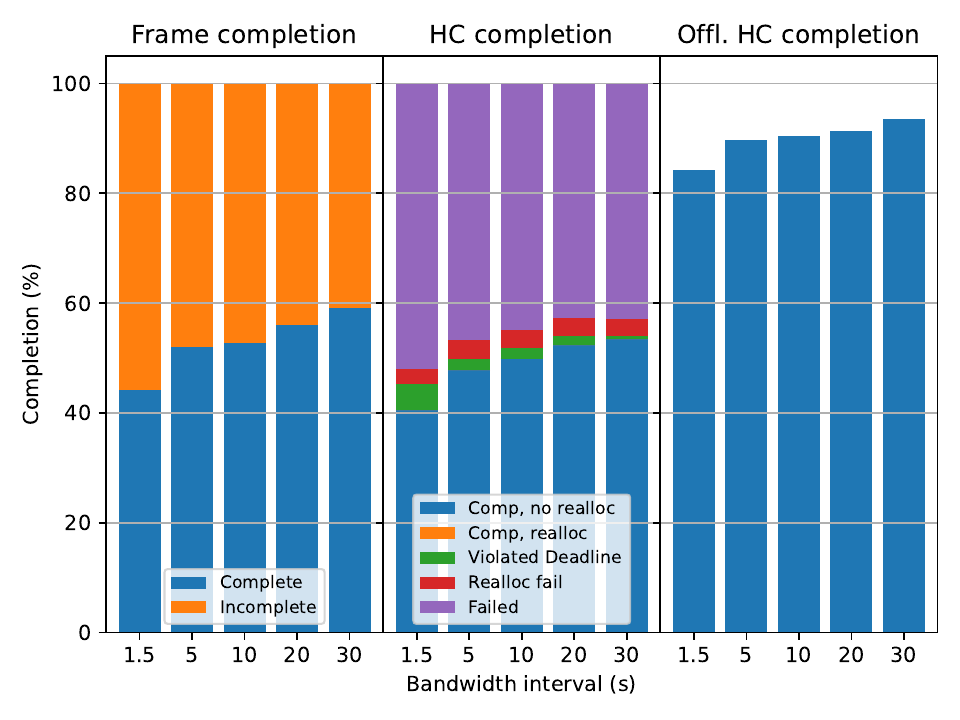}
    \caption{Bandwidth Interval Tests: Task completion across various categories.} 
    \label{fig:bandwidth_comp}
\end{figure}

We require a dynamic estimation of network throughput to properly schedule network use and implement the data structure of the discretised network link. To do so, we must perform active probes to all edge devices and build up an average of the network throughput, which will be used to adjust the EMWA of the estimated bandwidth. However, using an active approach such as this introduces a communication overhead. These overheads can negatively impact the transfer of task input, resulting in penalties in task completion, particularly around scenarios that involve task offloading.
To properly explore how much these overheads affect task completion, we perform a series of short tests using a 30 min slice of the weighted 4 scenario, while varying the frequency at which the bandwidth update mechanism is invoked as follows: 1.5s, 5s, 10s, 20s, and 30s. Other than the modifications to the bandwidth update frequency, the network setup remains the same. We can immediately observe from Fig. \ref{fig:bandwidth_comp}, that as we decrease the frequency in which we invoke the bandwidth estimation mechanism, the frame completion rate increases. The figure also shows that this increase in frame completion is correlated to an increase in the number of low-priority tasks completed as we decrease the frequency. Additionally, we can see that the number of tasks that violate their deadlines also decreases as we reduce the frequency with which the bandwidth mechanism is invoked. The increase in the total number of low-priority tasks completed without reallocation is related to two factors. First, when a bandwidth update test is performed, the network discretisation must be regenerated to reflect updated network conditions. While this data-structure updates, no tasks can be allocated, introducing delays into the internal job queue of the scheduler. Second, when these network tests are performed more frequently, it is more likely that they will occur alongside ongoing image transfer operations. This will cause the system to believe that the available network bandwidth is lower than it truly is, leading to more conservative network link representations. The impact of the additional congestion can be seen further in Fig. \ref{fig:bandwidth_comp}, where we observe that the completion rate of offloaded tasks steadily increases as we further reduce the frequency of bandwidth updates.

\subsection{Network Traffic Congestion Tests}
\begin{figure}
    \centering
    \includegraphics[width=\columnwidth]{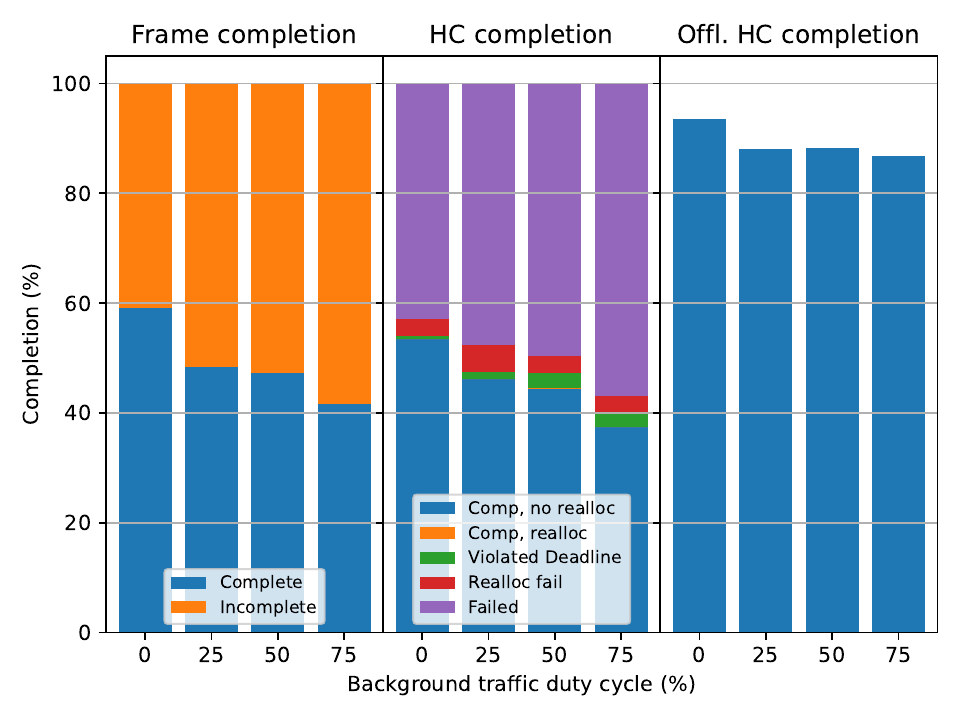}
    \caption{Network Traffic Test: Task completion across various categories.} 
    \label{fig:network_completion}
\end{figure}

Bursty background traffic can negatively affect the performance of the system. When these periods of burst occur after a bandwidth estimation test is performed, it results in a stale bandwidth estimate. This will induce inaccurate communication while scheduling the tasks and lead to task placement errors such as deadline violations. To understand the impact network congestion has on the performance task offloading we perform four 30 minute tests under the weighted 4 scenario with a traffic generator we implemented using the Packet\_MMAP API in Linux, to send 1024 bytes frames in bursts. We model it such that the bursts are set to a duty cycle of the bandwidth update interval (in these experiments we used a 30 second interval). We set the duty cycle across the four cycles to the following parameters: 0, 25, 50, 75. We observe in Fig. \ref{fig:network_completion} that as the active period increases, there is a decrease in the number of frames completed in the experiments. Between the absence of additional background traffic and traffic at 75\% of the period, performance drops 18\%. Overall, while there is a performance drop, Fig. \ref{fig:network_completion} shows that the bulk of this performance drop can be attributed to the failure to allocate tasks. Task deadline violations do not see a significant increase across the four scenarios; this may be because the dynamic bandwidth estimation mechanism correctly estimates that the increase in traffic makes allocating tasks less viable. We observe in Fig. \ref{fig:network_completion}, that the only significant drop in the performance of offloaded tasks is from 0\% background traffic to 25\%; while there is a minor drop between 50\% and 75\% overall performance remains consistent across the three background traffic scenarios. Indeed, as the available network throughput decreases, the time to transfer an image from one device to another increases, making offloading more costly. We can see from Table \ref{tab:core_allocation} that, as the window to allocate tasks decreases, the system attempts to compensate for this by allocating tasks a higher number of cores, so that they process faster. What is interesting is that, in network tests with an active cycle of 75\%, the overall performance of the system regarding frame completion and low-priority task completion is comparable one of the interval tests where the bandwidth update takes place every 1.5 seconds. This indicates that these higher bandwidth update values introduce a large amount of congestion to the network, and that this is not mitigated by more accurate bandwidth estimations.

\begin{table}[]
\resizebox{\columnwidth}{!}{%
\begin{tabular}{l|llll}
\textbf{Duty Cycle} & \textbf{0\%} & \textbf{25\%} & \textbf{50\%} & \textbf{75\%} \\ \hline
\textbf{Two Core}   & 100\%        & 95.97\%       & 96.58\%       & 87.70\%       \\ \hline
\textbf{Four Core}  & 0\%          & 4.03\%        & 3.42\%        & 12.30\%       \\ \hline
\end{tabular}%
}
\\
\caption{Network Traffic Test: Core allocation of successfully allocated tasks}
\label{tab:core_allocation}
\end{table}

\section{Conclusions}
\label{sec:conclusions}
In this paper we have proposed a lightweight scheduling model for both computational resources and the network discretisation. The model is paired with a dynamic bandwidth update mechanism. The overall reduction in latency of the above provides strong benefits under heavily loaded and congested network conditions, by reducing the amount of erroneous task placements the scheduler makes. It also allows for more frequent reallocations of preempted tasks under resource scarcity. However, under lightly loaded conditions, such improvements do not necessarily yield the same benefits, when compared to more accurate and thorough approaches. Additionally, the frequency of bandwidth estimation tests impacts the performance of the system, as performing them too frequently introduces additional network congestion and leads to underestimating network throughput.

In future work, we would like to explore a contextual multi-scheduler approach, utilising a more accurate approach under lightly loaded conditions and switching to light-weight scheduling abstraction models in times of higher network load. Additionally, we want to explore a mechanism for dynamically varying the frequency of bandwidth tests during run-time based on network congestion.

\label{sec:acknowledgements}
\section {Acknowledgements}
This publication has emanated from research conducted with the financial support of Taighde Éireann – Research Ireland under Grant numbers 18/CRT/6222 and 13/RC/2077\_P2. For the purpose of Open Access, the author has applied a CC BY public copyright licence to any Author Accepted Manuscript version arising from this submission.
\bibliographystyle{ieeetr}
\bibliography{bibliography}

\end{document}